\documentclass{article}
\usepackage[utf8]{inputenc}
\DeclareMathAlphabet{\mathpzc}{OT1}{pzc}{m}{it}

\usepackage{amsmath} 
\usepackage{amsfonts}
\usepackage{dsfont}
\usepackage[utf8]{inputenc}
\usepackage[english]{babel}
\usepackage[T1]{fontenc}
\usepackage{physics}
\usepackage{authblk}
\usepackage{tikz}
\usepackage{subcaption}
\usepackage{graphicx}
\newmuskip\pFqmuskip
\newcommand*\pFq[6][8]{%
  \begingroup 
  \pFqmuskip=#1mu\relax
  \mathcode`\,=\string"8000
  \begingroup\lccode`\~=`\,
  \lowercase{\endgroup\let~}\pFqcomma
  {}_{#2}F_{#3}{\left[\genfrac..{0pt}{}{#4}{#5};#6\right]}%
  \endgroup
}
\newcommand{\pFqcomma}{\mskip\pFqmuskip}

\makeatletter
\newsavebox{\@brx}
\newcommand{\llangle}[1][]{\savebox{\@brx}{\(\m@th{#1\langle}\)}%
  \mathopen{\copy\@brx\kern-0.5\wd\@brx\usebox{\@brx}}}
\newcommand{\rrangle}[1][]{\savebox{\@brx}{\(\m@th{#1\rangle}\)}%
  \mathclose{\copy\@brx\kern-0.5\wd\@brx\usebox{\@brx}}}
\makeatother

\begin{document}

\title{Entanglement of Free Fermions on Johnson Graphs}
\author[1]{Pierre-Antoine Bernard}
\author[2]{Nicolas Crampé}
\author[1]{Luc Vinet}
\affil[1]{Centre de recherches mathématiques, Université de Montréal, P.O. Box 6128, Centre-ville
Station, Montréal (Québec), H3C 3J7, Canada,}
\affil[2]{Institut Denis-Poisson CNRS/UMR 7013 - Université de Tours - Université d’Orléans, Parc de
Grandmont, 37200 Tours, France.}

\maketitle
\begin{abstract}
 Free fermions on Johnson graphs $J(n,k)$ are considered and the entanglement entropy of sets of neighborhoods is computed. For a subsystem composed of a single neighborhood, an analytical expression is provided by the decomposition in irreducible submodules of the Terwilliger algebra of $J(n,k)$ embedded in two copies of $\mathfrak{su}(2)$. For a subsytem composed of multiple neighborhoods, the construction of a block-tridiagonal operator which commutes with the entanglement Hamiltonian is presented, its usefulness in computing the entropy is stressed and the area law pre-factor is discussed.
\end{abstract}
\section{Introduction}

In quantum systems, observables attached to different regions are generally correlated to an extent that depends on the state, the geometry, etc. The notion of entanglement entropy quantifies the correlation between a subsystem and its complementary part. It plays a central role in many branches of quantum theory, notably in many-body physics \cite{Amico:2007ag, Latorre_2009, Peschel_2012}.

In recent papers, tools from the study of time and band limiting problems \cite{Landau1985,Slepian1983} and from the theory of association schemes \cite{brouwer2012distance,inbook} have been applied to the computation of this quantity. They were used for models of free fermions hopping on chains \cite{Cramp__2019,Cramp__2020} or on the vertices of distance-regular graphs. In the latter case, the Hadamard \cite{ crampe2020entanglement} and the Hamming graphs \cite{Hamming, jafarizadeh2014entanglement, Jafarizadeh2015EntanglementEI} were specifically studied and in some instances analytical expressions for the entanglement entropy and thermodynamic limits were obtained. Bethe ansatz techniques were also shown to be useful to study such problems \cite{bernard2020heun}. 

We here pursue this exploration and consider the entanglement of free fermions living on Johnson graphs $J(n,k)$. These graphs are well known to be distance-regular and to belong to a $P$- and $Q$- polynomial association scheme \cite{Bannai1984AlgebraicCI}. The adjacency and dual adjacency matrices of these graphs span an algebra referred to as the Terwilliger algebra $\mathcal{T}$ of the Johnson scheme \cite{thini,thinii,thiniii}. Most objects we shall use to compute the entropy arise from this structure. 

We shall take the system to be in its ground state. The entanglement entropy can be computed from the eigenvalues of the chopped correlation matrix $C$ \cite{Peschel_2003,Peschel_2009}. Obtaining the spectrum of this matrix is thus the main aim of this paper. This shall be done in particular for subsystems corresponding to neighborhoods. A neighborhood is the set of all the vertices at a given distance from a reference site. For such subsystems, the chopped correlation matrix $C$ is part of the Terwilliger algebra $\mathcal{T}$ of the Johnson scheme and we can decompose the vector space on which $C$ is acting in irreducible $\mathcal{T}$-submodules. This process which can be seen as breaking down the graph $J(n,k)$ into a direct sum of chains (or paths) greatly simplifies the diagonalization of $C$.

The determination of the irreducible $\mathcal{T}$-submodules was considered in \cite{LEVSTEIN20071621} and in \cite{GAO2014164,Obse,TAN2019157} with an approach based on the theory of Leonard pairs \cite{TERWILLIGER2003463}. These shall be obtained here using a different route. It is known that the Johnson graphs can be embedded in hypercubes \cite{geo}. Translated in algebraic terms, this statement implies that $\mathcal{T}$ can be embedded in two copies of the Terwilliger algebra of the hypercube. Since the decomposition in irreducible modules is known for the latter \cite{Hamming,GO2002399}, it will also yield the decomposition of $\mathcal{T}$. This perspective has the advantage of being related to the coupling of two $\mathfrak{su}(2)$ representations and establishes a relation between the Terwilliger algebra of the Johnson scheme and the Hahn algebra $\mathfrak{h}$ \cite{GRANOVSKII19921}.

While expressing the graph $J(n,k)$ as a sum of chains diagonalizes $C$ for subsystems made out of a single neighborhood, it is not sufficient for subsystems composed of multiple neighborhoods. To alleviate this issue, we shall construct a block-tridiagonal operator $T$ which shares with the chopped correlation matrix a set of common eigenvectors. It will be referred to as a generalized Heun operator \cite{GB_2018}. This procedure proves analogous to the introduction of a commuting second order differential operator in the study of time and band limiting \cite{Landau1985,Slepian1983}. This approach was used to compute the entropy of free fermions on graphs in \cite{crampe2020entanglement}.

The paper is divided in four parts. In section \ref{s1}, the Hamiltonian of free fermions on Johnson graphs is presented and diagonalized. The single-particle excitation energies are given and the ground state is defined. Entanglement entropy is discussed in section \ref{s2}. We describe the relation between the spectrum of the reduced density matrix and the spectrum of the chopped correlation matrix. In section \ref{s3}, we give an overview of the Terwilliger algebra of the Johnson scheme, we find the irreducible components of its standard module and we show that this decomposition diagonalizes $C$ for single-neighborhood subsystems. Entanglement entropies are computed and comments are made on the relation between $\mathcal{T}$ and $\mathfrak{h}$. In section \ref{s4}, the generalized Heun operator $T$ is constructed and the entanglement entropy of large bundles of neighborhoods is examined.

\section{Free fermions on Johnson graphs}
\label{s1}
The Johnson graph $J(n,k)$, $k \leq n/2$, is constructed in the following way. First, consider the subsets $x \subset \{1, 2, \dots, n\}$ of cardinality $k$ as the elements of the set of vertices $X$. Then, take two subsets to be connected by an edge when they differ only by one element. This yields a graph with $|X| = \binom{n}{k}$ vertices and a diameter of $k$. Given two subsets $x$ and $y$ in $X$, their distance is
\begin{align}
    d(x,y) = k - |x \cap y|.
\end{align}
 In this paper, we consider free fermions living on the vertices of Johnson graphs $J(n,k)$. In particular, we study fermionic systems for which the hopping constant $\alpha_{d(x,y)}$ between the sites $x$ and $y$ is real and depends only on $d(x,y)$. More precisely, the Hamiltonian is defined as
\begin{align}
    \widehat{\mathcal{H}} = \sum_{x,y \in X} \alpha_{d(x,y)}c_x^\dagger c_{y}, 
    \label{hamilt}
\end{align}
\noindent where $c_x^\dagger$ and $ c_{x}$ are creation and annihilation operators associated to the site $x$. They satisfy the following canonical relations:
\begin{align}
    \{c_x, c_{y}\} = 0, \quad \quad \{c_x^\dagger,c_{y}^\dagger\} = 0, \quad \quad  \{c_x,c_{y}^\dagger\} = \delta_{x y}, \quad \quad \forall x,y \in X.
    \label{ferm}
\end{align}
\noindent We note that this model contains $k + 1$ parameters and that $\alpha_0$ is related to the presence of an external magnetic field. We can also give an alternative expression for $\widehat{\mathcal{H}}$. Let $\text{Mat}_X(\mathbb{C})$ denote the space of matrices with complex entries and with rows and columns labeled by elements in $X$. For $i \in \{0, 1, \dots, k\}$, one defines the $i^{\text{th}}$ adjacency matrix $A_i$ of $J(n,k)$ as the matrix in $\text{Mat}_X(\mathbb{C})$ whose entry $[A_i]_{xy}$ is
\begin{align}
[A_i]_{xy} = 
\left\{
    \begin{array}{ll}
    	1  & \mbox{if } d(x,y) = i, \\
    	0 & \mbox{otherwise. }
    \end{array}
\right.
\label{adjazzz}
\end{align}
\noindent Each vertex $x \in X$ of the graph is represented by a column vector $\ket{x}$ which has a $1$ in the row $x$ as its unique non-zero entry. In terms of the vectors of operators $\hat{c}^\dagger = \sum_x c^\dagger_x\bra{x}$ and $\hat{c} = \sum_x \ket{x }c_x$, the Hamiltonian can be rewritten as 
\begin{align}
    \widehat{\mathcal{H}} = \hat{c}^\dagger\Big[ \sum_{i = 0}^{k} \alpha_i A_i \Big] \hat{c}.
    \label{hamil}
\end{align}
\subsection{Diagonalization and energies}
\label{secdia}
\noindent To diagonalize \eqref{hamil} it is sufficient to diagonalize $ \sum_{i = 0}^{k} \alpha_i A_i $. Since the Johnson graph $J(n,k)$ is distance-regular, a result from the theory of association scheme implies that $A_i$ can be expressed as a polynomial of degree $i$ in $A \equiv A_1$ \cite{Bannai1984AlgebraicCI}:
\begin{equation}
    \begin{split}
            A_i  &= (-1)^i \binom{k}{i}R_i(A + k;0,n-2k,k)\\
            &= (-1)^i \binom{k}{i} \sum_{r=0}^{i} \binom{i}{r} \frac{(k-r)!}{(r)!(k)!}\prod_{\ell = 0}^{r-1} (-A - k + \ell(n-2k+1) + \ell^2),
    \label{Hahn}
    \end{split}
\end{equation}
where $R_i$ refers to the dual Hahn polynomial of degree $i$ \cite{koekoek1996askey}. Thus, we only have to diagonalize $A$. Its spectrum is known to be\footnote{The spectrum of $A$ is usually presented as $(k-u)(n-k-u+1) - k$ with $u \in \{0,1,\dots,k\}$ \cite{Bannai1984AlgebraicCI}. The label $j = n/2 - u$ is used in \eqref{eigenj} instead for reasons we discuss in section \ref{irrep}.}: 
\begin{align}
    \theta_j =  j(j+1) - \frac{(n-2k)^2}{4} - \frac{n}{2}, \quad \text{with} \quad j \in \{\frac{n}{2} - k, \frac{n}{2} - k +1, \dots, \frac{n}{2}\}.
    \label{eigenj}
\end{align}
\noindent For now, we shall also refer to the basis vectors of the $j^{\text{th}}$ eigenspace of the adjacency matrix with $\ket{\theta_j,\ell}$:
\begin{align}
    A \ket{\theta_j, \ell } = \theta_j \ket{\theta_j,\ell}.
\end{align}
The label $\ell$ accounts for the degeneracy $D_j$ of the eigenspace $j$. The actual construction of these vectors and an explicit expression for $D_j$ is also discussed in section \ref{irrep}. For each $j$ and $\ell$, let us now define a new pair of creation and annihilation operators:
\begin{align}
    \bar{c}_{j,\ell}^\dagger  = \sum_{x \in X} \bra{x}\ket{\theta_j, \ell} c_x^\dagger \quad \quad \text{and} \quad \quad \bar{c}_{j,\ell}  =\sum_{x \in X} \bra{\theta_j, \ell }\ket{x} c_x
\end{align}
\noindent One can check that they satisfy the same canonical relations as $c_x$ and $c^\dagger_y$ and that they allow the diagonalization of $\widehat{\mathcal{H}}$:
\begin{align}
    \widehat{\mathcal{H}} = \sum_{j = \frac{n}{2} - k}^{\frac{n}{2}} \sum_{\ell = 1}^{D_j} \Omega_j  \bar{c}_{j,\ell}^\dagger\bar{c}_{j,\ell},
\end{align}
\noindent where
\begin{align}
    \Omega_j = \sum_{i = 0}^k \alpha_i (-1)^i\binom{k}{i}R_i(\theta_j + k ; 0, n-2k, k).
    \label{ener}
\end{align}
\noindent In some useful cases, formula \eqref{ener} reduces to a simplier form. For instance, a model restricted to nearest neighbor hopping is obtained by taking $\alpha_i = 0$ for $i > 1$. Then, $\Omega_j$ is proportional to $\theta_j$ and hence quadratic in $j$.  Furthermore, cases where the hopping terms decrease exponentially with the distance are modeled by taking $\alpha_i = e^{-ci}$ for all $i$, where $c \geq 0$. Then \eqref{ener} corresponds to the generating function of the dual Hahn polynomials and strictly grows with $j$:
\begin{align}
    \Omega_{j} = (1-e^{-c})^{\frac{n}{2}-j} \pFq{2}{1}{\frac{n}{2}-k-j,-\frac{n}{2} + k - j}{1}{e^{-c}},
\end{align}
where is ${}_2F_1$ is the hypergeometric function \cite{koekoek1996askey}. 
\subsection{Ground state}

\noindent Let $|0 \rrangle$ be the vacuum state annihilated by all the operators $\bar{c}_{j,\ell}$. In fermionic systems, the ground state $|\Psi_0 \rrangle$ is the state for which all the energy levels $\Omega_j < 0$ are occupied. We denote $SE$ the set of all the integers $j \in \{\frac{n}{2} - k, \frac{n}{2}-k+1, \dots, \frac{n}{2}\}$ associated to negative single-particle excitation energies $\Omega_j$. For fixed parameters $\alpha_i$, one can easily identify $SE$ by computing the values taken by \eqref{ener}. The ground state $|\Psi_0 \rrangle$ is then given by
\begin{align}
    |\Psi_0 \rrangle = \Big[\prod_{j\in SE} \ \prod_{\ell = 1}^{D_j}\bar{c}_{j,\ell}^\dagger \Big]  |0 \rrangle .
\end{align}
\noindent For nearest neighbor or exponentially decreasing hopping, $\Omega_j$ grows with $j$ and so $SE$ corresponds to a set $\{\frac{n}{2} - k, \frac{n}{2}-k+1, \dots,j_0\}$ for some integer or half integer $j_0$. As we shall see in section \ref{s2}, the information we need to compute entanglement entropies is contained in the correlation matrix $\widehat{C}$. It is the matrix whose components $\widehat{C}_{xy}$ are defined as 
\begin{align}
    \widehat{C}_{xy} = \llangle \Psi_0 | c_x^\dagger c_{y}|\Psi_0 \rrangle, \quad \quad \text{where }  x, y \in X.
\end{align}
\noindent We can use the eigenbasis of $A$ $\{\ket{\theta_j,\ell} : j \in \{\frac{n}{2}-k, \dots, \frac{n}{2}\}, \ell \in \{1, \dots, D_j\}\}$ to express $c_x^\dagger$ and $c_y$ in terms of $\Bar{c}_{j,\ell}^\dagger$ and $\Bar{c}_{j,\ell}$. Then, simple algebraic manipulations yields
\begin{equation}
\begin{split}
     \widehat{C} &= \sum_{j \in SE} \sum_{\ell = 1}^{D_j} \ket{\theta_j, \ell}\bra{\theta_j,\ell} \\
    &= \sum_{j \in SE} E_j \equiv \pi_{SE},
    \label{proen}
\end{split}
\end{equation}
\noindent where $E_j$ is the projection operator onto the eigenspace $j$ of $A$ and $\pi_{SE}$ is the projection operator onto all the eigenspaces associated to an integer or half integer in $SE$. 
\section{Entanglement entropy}
\label{s2}
The $i^{\text{th}}$ neighborhood with respect to a vertex $x_0$ is the set of sites $x \in X$ such that $d(x_0,x)=i$. The projector onto this set of vertices is
\begin{align}
    E^*_i = \sum_{\substack{x \in X \\ d(x_0,x) = i}} \ket{x}\bra{x}.
    \label{dp}
\end{align}
Let us take a subset $SD$ of distances in $\{0,1,\dots,k\}$. We refer to the bundle of neighborhoods of $x_0$ associated to integers in $SD$ as the subsytem $1$ or $SV \subset X$. The projection operator onto this subsystem is
\begin{align}
\begin{split}
     \pi_{SV} &= \sum_{x \in SV} \ket{x}\bra{x} = \sum_{i \in SD} E^*_i .
    \label{propoz}
\end{split}
\end{align} 
Similarly, we refer to its complement $X\backslash SV$ as the subsystem $2$. In the ground state, the reduced density matrix $\rho_1$ and the von Neumann entropy $S$ of the subsystem $1$ are respectively defined as
\begin{align}
    \rho_1 = \text{tr}_2 |\Psi_0 \rrangle \llangle \Psi_0 | \quad \quad \text{and} \quad \quad  S = - \text{tr}(\rho_1 \ln{\rho_1}).
    \label{vN}
\end{align}
\noindent The entanglement entropy $S$ measures to which extent the state of $SV$ is correlated with the state of $X\backslash SV$. Once the eigenvalues of $\rho_1$ are determined, computing $S$ is immediate. It is known that these eigenvalues are related to those of the chopped correlation matrix $C$ \cite{Peschel_2003,Peschel_2009}, which is defined as
\begin{align}
    C = |\widehat{C}_{xy}|_{xy \in SV},
\end{align}
\noindent and is given by
\begin{align}
    C = \pi_{SV}  \pi_{SE}  \pi_{SV}
    \label{c1}
\end{align}
\noindent in terms of the projection operators \eqref{proen} and \eqref{propoz}. The relation between the spectra of $\rho_1$ and $C$ allows to rewrite \eqref{vN} in terms of the eigenvalues $\lambda$ of the chopped correlation matrix and their degeneracy $\mathcal{D}_\lambda$ \cite{Carrasco_2017}:
\begin{align}
    S = - \sum_{\lambda} \mathcal{D}_\lambda \left[\lambda \ln{(\lambda)} + (1-\lambda)\ln{(1-\lambda)}\right].
    \label{entropy}
\end{align}
\noindent Thus, we are interested in diagonalizing $C$. By developing expression \eqref{c1}, we find
\begin{align}
    C = \sum_{i,i' \in SD} \sum_{j \in SE} E^*_i E_j E^*_{i'}.
    \label{c2}
\end{align}
Since $J(n,k)$ is distance-regular, we know from the theory of association schemes that the set of projectors onto the eigenspaces of $A$, i.e. 
\begin{align}
    \{E_{\frac{n}{2} - k},E_{\frac{n}{2} - k + 1}, \dots, E_{\frac{n}{2}}\},
\end{align}
\noindent and the set of projectors onto neighborhoods, i.e.
\begin{align}
    \{E_{0}^*, E_{1}^*, \dots, E_{k}^*\},
\end{align}
generate an algebra referred to as the Terwilliger algebra $\mathcal{T}$ of the Johnson scheme  \cite{thini,thinii,thiniii}. From formula \eqref{c2}, we see that the chopped correlation matrix is the representation of an element in the algebra $\mathcal{T}$. Thus, decomposing the vector space $\mathbb{C}^{|X|}$ on which $C$ is acting in its irreducible $\mathcal{T}$-submodules simplifies the diagonalization by allowing to work on one submodule at a time. Obtaining this decomposition is the aim of the next section.

\section{The Terwilliger algebra of the Johnson scheme}
\label{s3}
First, we present an overview of $\mathcal{T}$, the algebra spanned by the primitive idempotents $E_i$ and dual primitive idempotents $E^*_i$. This algebra is also generated by the set of adjacency matrices $\{A_0, A_1, \dots, A_k\}$ defined by \eqref{adjazzz} and the set of dual adjacency matrices $\{A^*_{0}, A^*_{1}, \dots, A^*_k\}$. These are diagonal matrices in $ \text{Mat}_X(\mathbb{C})$ for which the non-zero entries are given by
\begin{align}
    [A_i^*(x_0)]_{xx} = \binom{d}{k} [E_{\frac{n}{2} - i}]_{x_0 x},
    \label{defdual}
\end{align}
\noindent where $x_0$ is the reference vertex in \eqref{dp}. Note, that we use the simplified notation: $A^*_i = A^*_i(x_0)$ and $E^*_i = E^*_i(x_0)$. By construction, the set of vectors $\{\ket{x}\}_{x \in X}$ associated to sites in the graph gives an eigenbasis of the dual adjacency matrices. In particular, it is known \cite{GAO2014164, thiniii} that \eqref{defdual} implies
\begin{align}
    A^* \ket{x} = \left(n - 1 - \frac{n(n-1)}{k(n-k)}d(x_0,x) \right)\ket{x}.
    \label{actiona}
\end{align}
 Recall that \eqref{Hahn} gives $A_i$ as a polynomial of $A$. Since $J(n,k)$ is Q- polynomial, we also have that its $i^{\text{th}}$ dual adjacency matrix $A^*_i$ is a polynomials of degree $i$ in $A^* \equiv A_1^*$ and thus
\begin{align}
    \mathcal{T} = \langle A, A^* \rangle.
\end{align}

The commuting algebra spanned only by the adjacency matrices $A_i$ (or equivalently by the projectors $E_i$) of an association scheme is referred to as its Bose-Mesner algebra. The elements in the Bose-Mesner algebra of $J(n,k)$ verify \cite{Bannai1984AlgebraicCI,brouwer2012distance}:  
\begin{itemize}
        \item $A_0 = \mathds{1}_{|X| \times |X|}$ and $E_{\frac{n}{2}} = \frac{J_{|X| \times |X|}}{|X|}$ ;
        \item $\sum_{i =0}^k A_i  = J_{|X| \times |X|}$ and $\sum_{j = \frac{n}{2} - k}^{\frac{n}{2}} E_j  = \mathds{1}_{|X| \times |X|}$ ;
        \item $A_{i_1} \circ A_{i_2} = \delta_{i_1 i_2} A_{i_1}$ and $E_{j_1} E_{j_2} = \delta_{j_1j_2}E_{j_1 j_2}$ ; 
        \item $A_{i_1} A_{i_2} = \sum_{i_3=0}^k p_{i_1 i_2}^{i_3} A_{i_3}$ and $E_{j_1} \circ E_{j_2} = \frac{1}{|X|}\sum_{j_3=\frac{n}{2}-k}^{\frac{n}{2}} q_{\frac{n}{2} - j_1, \frac{n}{2} - j_2}^{\frac{n}{2} -j_3} E_{j_3}$,
    \end{itemize}
\noindent where $(A\circ B)_{mn} = A_{mn}B_{mn} $ is the entry-wise product, $J$ is the matrix of ones and $p_{i_1 i_2}^{i_3}$ and $q_{i_1 i_2}^{i_3}$ are real coefficients. Furthermore, there exist some coefficients $p_{i}(j)$ and $q_{\frac{n}{2}-j}(i)$ such that
\begin{align}
    A_i = \sum_j p_i(j) E_j \quad \quad \text{and} \quad \quad E_j = \frac{1}{|X|}\sum_ i q_{\frac{d}{2} - j}(i) A_i.
\end{align}
\noindent The connection of the Johnson scheme with the Hahn polynomials and dual Hahn polynomials exploited before in \eqref{Hahn} stems from this Bose-Mesner algebra \cite{Bannai1984AlgebraicCI}. 

The commuting algebra spanned by the dual adjacency matrices is the dual Bose-Mesner algebra of the scheme. The relations verified by its generators are:  
\begin{itemize}
        \item $\sum_{i = 0}^{k} E_i^*  = \mathds{1}_{|X| \times |X|}$;
        \item $A^*_{i_1} A^*_{i_2} = \sum_{i_3=0}^k q_{i_1 i_2}^{i_3} A^*_{i_3}$ and $E_{i_1}^* E_{i_2}^* = \delta_{i_1 i_2}E_{i_1}^* $ ;
        \item  $A^*_{i_1} = \sum_{i_2} q_{i_1}({i_2}) E^*_{i_2}$ and $E^*_{i_1} = \frac{1}{|X|}\sum_{i_2} p_{i_1}({i_2}) A^*_{i_2}$.
\end{itemize}

 The Johnson graphs can be embedded in hypercubes \cite{geo}. As we shall see, this can be used to embed the Terwilliger algebra of the Johnson scheme $\mathcal{T}$ in two copies of the Terwilliger algebra of the hypercube $\mathds{T}$. Since the decomposition in irreducible modules is known for $\mathds{T}$, we can use this relation to obtain the equivalent decomposition for $\mathcal{T}$.
 
\subsection{Embedding of $J(n,k)$ in $H(n,2)$}
\label{secemb}
The hypercube graph $H(n,2)$ is distance-regular and a special case of a Hamming graph $H(n,q)$. Its vertices are all the binary tuples of length $n$ composed of zeros and ones. Two vertices $v = (v_1,v_2,\dots,v_n)$ and $v' = (v_1',v_2',\dots,v_n')$ are connected by an edge if there exists a unique position $i \in \{1, 2,\dots,n\}$ such that $v_i \neq v_i'$. The distance between any pair of sites $v$ and $v'$ is given by their Hamming distance $\partial(v,v')$:
\begin{align}
    \partial(v,v') = \# \{i \in \{1, \dots,n\}\ :\  v_i \neq v_i'\}.
\end{align}
\noindent Each vertex $v = (v_1, \dots, v_n)$ in $H(n,2)$ can be represented by a vector $\ket{v}$ in $(\mathbb{C}^{2})^{\otimes n}$ :
\begin{align}
    \ket{v} = \ket{v_1} \otimes \ket{v_2} \otimes \dots \otimes \ket{v_n},
    \label{basei}
\end{align}
\noindent where $\ket{0} = \binom{1}{0}$ and $\ket{1} = \binom{0}{1}$ are column vectors. In this basis, the first adjacency matrix  $\mathds{A}$ of $H(n,2)$ is
\begin{align}
    \mathds{A} = \sum_{i = 1}^n \underbrace{\mathds{1} \otimes \dots \otimes \mathds{1}}_{i-1 \text{ times}} \otimes \ \sigma_x \otimes \mathds{1} \otimes \dots \otimes \mathds{1},
    \label{adj1}
\end{align}
\noindent where $\sigma_x$ is the usual Pauli matrix. Indeed, one can check using \eqref{basei} and \eqref{adj1} that $\bra{v} \mathds{A} \ket{v'}$ is non-zero only when $v$ and $v'$ are neighbors in $H(n,2)$. Since $H(n,2)$ is distance-regular, it has its own Terwilliger algebra which we refer to as $\mathds{T}$. Similar to $\mathcal{T}$, it is generated by the first adjacency matrix $\mathds{A}$ and the first dual adjacency matrix $\mathds{A}^*$ of the $n$-cube:
\begin{align}
    \mathds{T} = \langle \mathds{A}, \mathds{A}^* \rangle.
\end{align}
$\mathds{A}^*$ is defined with respect to a reference vertex $v_0 = (0,0,\dots,0)$ in $H(n,2)$ through a relation similar to \eqref{defdual}. Its action on vectors $\ket{v}$ associated to sites in the hypercube is known to be given by \cite{GO2002399,thiniii}:
\begin{align}
    \mathds{A}^* \ket{v} = (n - 2 \partial(v_0,v)) \ket{v}.
    \label{actiono}
\end{align}
\noindent In the basis \eqref{basei}, on can check that it can be expressed as
\begin{align}
   \mathds{A}^* = \sum_{i = 1}^n \underbrace{\mathds{1} \otimes \dots \otimes \mathds{1}}_{i-1 \text{ times}} \otimes \ \sigma_z \otimes \mathds{1} \otimes \dots \otimes \mathds{1}.
   \label{adj2}
\end{align}
In the following, we shall refer to the projector onto the $i^{\text{th}}$ eigenspace of $\mathds{A}$ as $\mathds{E}_i$ and the projector onto the $i^{\text{th}}$ eigenspace of $\mathds{A^*}$ (i.e. the $i^{\text{th}}$ neighborhood of $v_0$) as $\mathds{E}^*_i$.

\begin{figure}[h]
    \centering
    \begin{tikzpicture}
\draw (-5,0) -- (-3,2);
\draw (-5,0) -- (-3,0);
\draw (-5,0) -- (-3,-2);

\draw (1,0) -- (-1,2);
\draw (1,0) -- (-1,0);
\draw (1,0) -- (-1,-2);

\draw (-3,2) -- (-1,2);
\draw (-3,2) -- (-1,0);

\draw (-3,-2) -- (-1,0);
\draw (-3,-2) -- (-1,-2);

\draw (-3,0) -- (-1,2);
\draw (-3,0) -- (-1,-2);

\draw[line width=0.4mm, black,dotted]   (-5,0) to[out= 40,in= 180] (-1,2);
\draw[line width=0.4mm, black,dotted]    (-1,2) to[out=-40,in=40] (-1,-2);
\draw[line width=0.4mm, black, dotted]    (-5,0) to[out=-40,in=180] (-1,-2);

\draw[black,fill=black] (-1,2) circle (0.15cm);
\node[] at (-1,2.4) {010};
\draw[black,fill=white] (-1,0) circle (0.15cm);
\node[] at (-1,-2.4) {001};
\draw[black,fill=black] (-1,-2) circle (0.15cm);
\node[] at (-5.5,0) {100};

\draw[black,fill=white] (-3,2) circle (0.15cm);
\node[] at (-3,0.4) {000};
\draw[black,fill=lightgray] (-3,0) circle (0.15cm);
\node[] at (-1,-0.4) {111};
\draw[black,fill=white] (-3,-2) circle (0.15cm);
\node[] at (-3,-2.4) {101};

\draw[black,fill=white] (1,0) circle (0.15cm);
\node[] at (-3,2.4) {110};
\draw[black,fill=black] (-5,0) circle (0.15cm);
\node[] at (1,0.5) {011};

\draw[black,fill=black] (2.5,0) circle (0.15cm);
\node[] at (2,0) {$\{1\} $};
\draw[black,fill=black] (5.44,-1.7) circle (0.15cm);
\node[] at (5.44,-2.12) {$\{3\}$};
\draw[black,fill=black] (5.44,1.7) circle (0.15cm);
\node[] at (5.44,2.12) {$\{2\} $};

\node[] at (-2,-3) {$H(3,2)$};
\node[] at (3.5,-3) {$J(3,1)$};

\draw[line width=0.4mm, black,dotted]   (2.5,0) -- (5.44,-1.7);
\draw[line width=0.4mm, black,dotted]    (5.44,-1.7) -- (5.44,1.7);
\draw[line width=0.4mm, black, dotted]    (5.44,1.7) -- (2.5,0);

\end{tikzpicture}

    \caption{An embedding of $J(3,1)$ in $H(3,2)$. The left figure represents the 3-cube. The vertex in gray is the reference vertex $v_0$ and the black vertices correspond to its first neighborhood. The dotted lines show which pairs are at a Hamming distance $\partial$ of two. The right figure represents the Johnson graph $J(3,1)$ with dotted lines connecting vertices at distance one. }
    \label{fig:subham}
\end{figure}
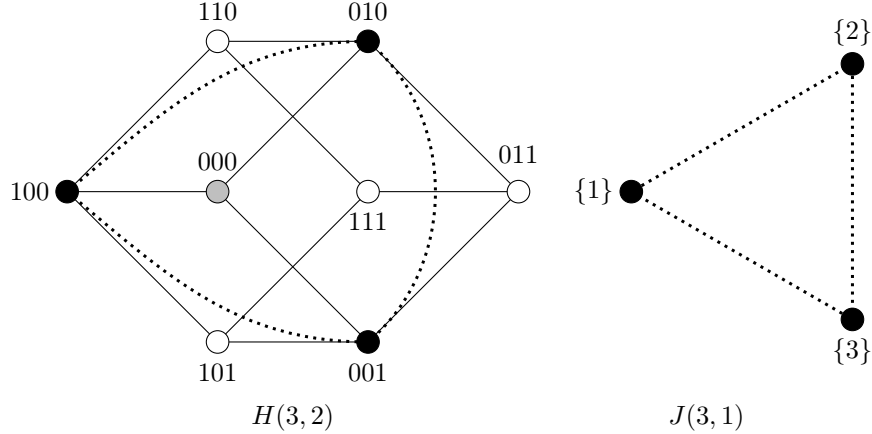

To embed the Johnson graph $J(n,k)$ in the hypercube $H(n,2)$, one has to map the $k$-subsets of $ \{1, 2, \dots, n\}$ onto the binary tuples $v = (v_1, v_2, \dots, v_n)$ of length $n$ containing $k$ ones. It can be achieved by associating the $k$-subset $x$ to the tuple $v(x)$ whose entry $v_i$ is
\begin{align}
    v_i = 
         \left\{
    \begin{array}{ll}
    	 1 & \mbox{if } i \in x,  \\
    	0 & \mbox{otherwise. }
    \end{array}
\right. 
\label{mapi}
\end{align}
\noindent For instance, \eqref{mapi} maps the vertex $x = \{3\}$ of $J(3,1)$ onto the tuple $v(x) = (0,0,1)$ of $H(3,2)$ (see Figure \ref{fig:subham}).

\noindent The vertices of $J(n,k)$ are thus identified with the sites in the $n$-cube which are in the $k^{\text{th}}$ neighborhood of $v_0$ , i.e. with the set of sites in $H(n,2)$ onto which the operator $\mathds{E}^*_k$ projects. Moreover, we see that two subsets are connected by an edge in $J(n,k)$ if and only if their Hamming distance in $H(n,2)$ is two. In other words, if $v(x)$ and $v(y)$ are the binary tuples associated to the subsets $x$ and $y$, we have
\begin{align}
    d(x,y) = \frac{1}{2} \partial(v(x),v(y)).
    \label{reldi}
\end{align}

This embedding can be translated in algebraic terms. Indeed, the first adjacency matrix of the Terwilliger algebra of the Johnson scheme $A$ corresponds to the restriction of the second adjacency matrix $\mathds{A}_2$ of $H(n,2)$ to the vertices in the $k^{\text{th}}$ neighborhood of $v_0$:
\begin{equation}
    \begin{split}
         A &= \mathds{E}_k^* \mathds{A}_2 \mathds{E}_k^*.
    \end{split}
    \label{Ajini}
\end{equation}
\noindent Since the matrices $\mathds{A}_i$ can be given in terms of Krawtchouk polynomials of degree $i$ in $\mathds{A}$ \cite{Bannai1984AlgebraicCI,koekoek1996askey}, we also have
\begin{equation}
    \begin{split}
         A = \frac{1}{2}\mathds{E}_k^* (\mathds{A}^2 - n) \mathds{E}_k^*.
    \end{split}
    \label{Aj}
\end{equation}
We now consider the relation between $A^*(x_0) = A^*$ and the generators in $\mathds{T}$. Without loss of generality, let us pick 
\begin{align}
    x_0 = \{n-k+1,n-k+2,\dots,n\}
\end{align}
as the reference vertex of $J(n,k)$. We recall that the action of $A^*$ and the action of $\mathds{A}^*$ on a vector $\ket{v(x)}$ is diagonal and given by \eqref{actiona} and \eqref{actiono} respectively. While the eigenvalue of $A^*$ on $\ket{v(x)}$ depends linearly on the distance $d(x_0,x)$, the eigenvalue of $\mathds{A}^*$ on $\ket{v(x)}$ depends linearly on the distance $\partial(v_0, v(x))$, with $v_0 \neq v(x_0)$ since these tuples are not in the same neighborhood of the $n$-cube. So, we define the following automorphism of the hypercube $H(n,2)$:
\begin{align}
    R = \underbrace{\mathds{1} \otimes \dots \otimes \mathds{1}}_{n-k \text{ times}} \otimes \underbrace{ \sigma_x \otimes \sigma_x \otimes \dots \otimes \sigma_x}_{k \text{ times}},
    \label{autoR}
\end{align}
which exchanges $v_0 = (0, 0,\dots,0)$ and 
\begin{align}
    v(x_0) = (\underbrace{0, 0, \dots, 0}_{n-k \text{ times}}, \underbrace{1, 1, \dots, 1}_{k \text{ times}})
\end{align}
 while preserving the distance between any pair of vertices. Formula \eqref{reldi} and \eqref{actiono} hence implies that
\begin{align}
    R  \mathds{A}^* R \ket{v(x)} =( n - 4 d(x_0,x)) \ket{v(x)}
\end{align}
\noindent and that $A^*$ corresponds to $R\mathds{A}^*R$ up to an affine transformation. Comparing their spectrum, we find
\begin{align}
  A^* = - \frac{(n-1)(n-2k)^2}{4k(n-k)} + \frac{n(n-1)}{4k(n-k)} R  \mathds{A}^* R .
\end{align}
\noindent Now let $\mathds{A}_{2^{n'}\times2^{n'}}$ and $\mathds{A}_{2^{n'}\times2^{n'}}^*$ refer to the adjacency and dual adjacency matrices of the hypercube $H(n',2)$. Similarly, let $\mathds{T}_{2^{n'}\times2^{n'}}$ refer to the Terwilliger algebra they span. Using \eqref{adj2} and \eqref{autoR}, one can check that
\begin{align*}
    R \mathds{A}_{2^{n}\times2^{n}}^* R = \mathds{A}_{2^{n-k}\times2^{n-k}}^* \otimes \mathds{1}_{2^{k}\times2^{k}} - \mathds{1}_{2^{n-k}\times2^{n-k}}\otimes\mathds{A}_{2^{k}\times2^{k}}^*.
\end{align*}
\noindent Thus, $A^*$ is contained in $\mathds{T}_{2^{n-k}\times2^{n-k}} \times \mathds{T}_{2^{k}\times2^{k}}$. Moreover, since
\begin{align}
      \mathds{A}_{2^{n}\times2^{n}} = \mathds{A}_{2^{n-k}\times2^{n-k}} \otimes \mathds{1}_{2^{k}\times2^{k}} + \mathds{1}_{2^{n-k}\times2^{n-k}}\otimes\mathds{A}_{2^{k}\times2^{k}},
\end{align}
\begin{align}
      \mathds{A}_{2^{n}\times2^{n}}^* = \mathds{A}_{2^{n-k}\times2^{n-k}}^* \otimes \mathds{1}_{2^{k}\times2^{k}} + \mathds{1}_{2^{n-k}\times2^{n-k}}\otimes\mathds{A}_{2^{k}\times2^{k}}^*,
\end{align}
\noindent and since the $\mathds{E}^*_k$ are polynomials of $\mathds{A}^*$, \eqref{Aj} guarantees that $A$ is also part of $\mathds{T}_{2^{n-k}\times2^{n-k}} \times \mathds{T}_{2^{k}\times2^{k}}$. Thus, $\mathcal{T}$ is embedded in the direct product of two copies of the Terwilliger algebra of the hypercube.

\subsection{ The irreducible $\mathcal{T}$-submodules and $\mathfrak{su}(2)$}
\label{irrep}
The generators of $\mathds{T}$ give a representation of the Lie algebra $\mathfrak{su}(2)$ \cite{Hamming, inbook}. Indeed, one can use \eqref{adj1} and \eqref{adj2} to show that the generators
\begin{align}
    {j^x} = \frac{\mathds{A}}{2}, \quad {j^z} = \frac{\mathds{A}^*}{2} \quad \text{and} \quad {j^y} = \frac{[{j^x}, {j^z}]}{2i}
    \label{su2}
\end{align}
\noindent obey the defining relations of $\mathfrak{su}(2)$. In particular, this representation corresponds to the $n$-fold product of the fundamental representation. For $H(n,2)$ with $n$ even (resp. odd) and for each $j$ in $\{ 0, 1,\dots, \frac{n}{2}\}$ (resp. in $\{ 1/2, 3/2, \dots, \frac{n}{2}\}$), the standard Clebsh-Gordan decomposition yields $\frac{2j + 1}{n+1}\binom{n+1}{\frac{n}{2} - j}$ orthogonal subspaces spanned by vectors $\{\ket{j,m}_{\ell}\}_{-j \leq m \leq j}$ such that
\begin{equation}
    \begin{split}
            \mathds{A} \ket{j,m}_\ell = 2j^x \ket{j,m}_\ell
            &= \sqrt{(j+m +1)(j-m)} \ket{j,m+1}_\ell  \\ & \quad + \sqrt{(j-m +1)(j+m)} \ket{j,m-1}_\ell
    \end{split}
\end{equation}
\noindent and
\begin{equation}
    \begin{split}
            \mathds{A}^* \ket{j,m}_\ell  
            = 2j^z \ket{j,m}_\ell =2m \ket{j,m}_\ell,
    \end{split}
\end{equation}
\noindent where the label $\ell \in \{1, \dots, \frac{2j + 1}{n+1}\binom{n+1}{\frac{n}{2} - j}\}$ indicates in which subspace of dimension $2j+1$ these vectors are contained. Since the vectors $\ket{j,m}_\ell$ are eigenvectors of $\mathds{A}^*$ of eigenvalue $2m$, they are in the $(\frac{n}{2} - m)^{\text{th}}$ neighborhood of $v_0 = (0,0,\dots,0)$, i.e.
\begin{align}
    \ket{ j, m}_\ell \in \text{span}\{\ket{v} : \partial(v_0,v) = \frac{n}{2} - m\}.
\end{align}
The embedding discussed in subsection \ref{secemb} allows to apply the relation between $\mathds{T}$ and $\mathfrak{su}(2)$ to the Terwilliger algebra of the Johnson scheme and to obtain the decomposition of $\mathbb{C}^{|X|}$ in irreducible $\mathcal{T}$-submodules. In terms of the ladder operators
\begin{align}
    j^+ = j^x + ij^y = \sum_{i=0}^{n-1} \mathds{E}_{i+1}^*\mathds{A} \mathds{E}^*_{i}
\end{align}
and
\begin{align}
    j^- = j^x - ij^y = \sum_{i=0}^{n-1} \mathds{E}_{i}^*\mathds{A} \mathds{E}^*_{i+1},
\end{align}
one can check using \eqref{Aj} that $A$ is given by
\begin{equation}
    \begin{split}
        A &= \frac{1}{2}\mathds{E}_k^*(\{j^+,j^-\} -n)\mathds{E}_k^*.
    \end{split}
\end{equation}
\noindent We also have the following relation:
\begin{align}
    \{j^+,j^-\} = 2\textbf{j}^2 - 2(j^z)^2,
\end{align}
\noindent where $\textbf{j}^2 = j_x^2 + j_y^2 + j_z^2$ is the Casimir operator of $\mathfrak{su}(2)$. Since $j^z = \frac{n}{2} - k$ on the $k^{\text{th}}$ neighborhood and since $[\mathds{E}_k^*, \textbf{j}^2] = 0$, we can rewrite the adjacency matrix $A$ of $J(n,k)$ as 
\begin{equation}
    \begin{split}
        A &= \mathds{E}_k^*\left(\textbf{j}^2 - \frac{(n-2k)^2}{4} - \frac{n}{2}\right).
        \label{Adjf}
    \end{split}
\end{equation}
\noindent So the first adjacency matrix of $J(n,k)$ is the restriction of the total Casimir operator of $\mathfrak{su}(2)$ to a single neighborhood of $H(n,2)$. For $A^*$, we note that the representation of $\mathfrak{su}(2)$ defined in \eqref{su2} corresponds to the coproduct of representations of lower dimension. For instance, we have $j^z =j^z_1 + j^z_2$ where
\begin{align}
    j^z_{1} =  \frac{\mathds{A}^*_{2^{n-k}\times 2^{n-k}}}{2} \otimes \mathds{1}_{2^{k}\times 2^{k}} \quad \quad \text{and} \quad \quad  j^z_{2} = \mathds{1}_{2^{n-k}\times 2^{n-k}} \otimes  \frac{\mathds{A}^*_{2^{k}\times 2^{k}}}{2}.
\end{align}
\noindent This allows to express the dual adjacency matrix of $J(n,k)$ as
\begin{align}
  A^* = - \frac{(n-1)(n-2k)^2}{4k(n-k)} + \frac{n(n-1)}{2k(n-k)} (j_1^z - j^z_2)
  \label{Adjff}
\end{align}
\noindent and implies that both generators of $\mathcal{T}$ are representing elements in $\mathfrak{su}(2) \otimes \mathfrak{su}(2)$. We can use \eqref{Adjf} and \eqref{Adjff} to derive expressions for the eigenvectors of $A$ and $A^*$. Let $\ket{j_1, m_1}_{\ell_1}$ be an eigenvector of $j_1^z$ in an irreducible $\mathds{T}_{2^{n-k}\times 2^{n-k}}$-submodule and let $\ket{j_2, m_2}_{\ell_2}$ be an eigenvector of $j_2^z$ in an irreducible $\mathds{T}_{2^{k}\times 2^{k}}$-submodule. When $m_1 + m_2 = \frac{n}{2} - k$, we see that
\begin{align}
    A^* \ket{j_1, m_1}_{\ell_1} \otimes \ket{j_2, m_2}_{\ell_2} = \theta^*_{m_1,m_2} \ket{j_1, m_1}_{\ell_1} \otimes \ket{j_2, m_2}_{\ell_2},
\end{align}
\noindent where
\begin{align}
    \theta_{m_1, m_2}^* = - \frac{(n-1)(n-2k)^2}{4k(n-k)} + \frac{n(n-1)}{2k(n-k)} (m_1 - m_2).
\end{align}
\noindent Since the vectors of this form are orthogonal and generate $\mathbb{C}^{|X|}$, they give an eigenbasis of $A^*$. By construction, we also note that $\ket{j_1, m_1}_{\ell_1} \otimes \ket{j_2, m_2}_{\ell_2}$ is in the $(\frac{n-k}{2} - m_1)-$th neighborhood of the Johnson graph, i.e.
\begin{align}
    \ket{j_1, m_1}_{\ell_1} \otimes \ket{j_2, m_2}_{\ell_2} \in \text{span}\{\ket{x} : d(x_0,x) = \frac{n-k}{2} - m_1 = \frac{k}{2} + m_2 \}.
\end{align}
Next, we can consider the eigenvectors of $A$. We can define the following subspaces $ V_{j_1, \ell_1, j_2, \ell_2}$ of $\mathbb{C}^{|X|}$:
\begin{align}
     V_{j_1, \ell_1, j_2, \ell_2} = \text{span}\{\ket{j_1, m_1}_{\ell_1} \otimes \ket{j_2, m_2}_{\ell_2} : m_1 + m_2 = \frac{n}{2} - k\}.
\end{align}
\noindent Since they are isomorphic for different values of $\ell_1$ and $\ell_2$, we also use the notation $V_{j_1, j_2} = V_{j_1, \ell_1, j_2, \ell_2}$. The vectors $\ket{j_1, m_1}_{\ell_1} \otimes \ket{j_2, m_2}_{\ell_2}$ diagonalize the operators $j_1^z$, $j_2^z$, $\textbf{j}^2_1$ and $\textbf{j}^2_2$. We know from the theory of angular momentum coupling how to construct an alternative basis $\{\ket{j, m } : |j_1 - j_2| \leq j \leq j_1 + j_2, \ m = \frac{n}{2} - k\}$ of $V_{j_1, \ell_1, j_2, \ell_2}$ diagonalizing $j^z$ and $\textbf{j}^2$ instead:
\begin{align}
    j_z \ket{j, n/2 - k} = (n/2 - k) \ket{j, n/2 -k}
\end{align}
\noindent and
\begin{align}
    \textbf{j}^2 \ket{j, n/2 - k} = j(j+1) \ket{j, n/2 - k}.
\end{align}
\noindent From expression \eqref{Adjf}, we deduce that these vectors diagonalize $A$:
\begin{align}
    A \ket{j, n/2 - k} = \left(j(j+1) - \frac{(n-2k)^2}{4} - \frac{n}{2}\right) \ket{j, n/2 - k}.
\end{align}
\noindent In particular, $\ket{j, n/2 - k} \in V_{j_1, \ell_1, j_2, \ell_2}$ is in the eigenspace $j$ of the adjacency matrix:
\begin{align}
    E_{j'} \ket{j, n/2 - k } = \delta_{j j'}\ket{j, n/2 - k },
\end{align}
\noindent and gives an explicit construction for the vectors $\ket{ \theta_j,\ell}$ of subsection \ref{secdia}. Now that we have a basis for the eigenspace $j$ of $A$, we see that its degeneracy $D_j$ is given by the number of subspaces $V_{j_1, \ell_1, j_2, \ell_2}$ such that $|j_1 - j_2| \leq j \leq j_1 + j_2$:
\begin{equation}
    \begin{split}
         D_j &= \sum_{\substack{j_1, j_2 \\ |j_1 - j_2| \leq j \\ j \leq j_1 + j_2}} D_{j_1, j_2} \\ &=  \sum_{\substack{j_1, j_2 \\ |j_1 - j_2| \leq j\\ j \leq j_1 + j_2 }} \frac{(2j_1 + 1)(2j_2 + 1)}{(n-k+1)(k+1)}\binom{n-k+1}{\frac{n-k}{2} - j_1}\binom{k+1}{\frac{k}{2} - j_2},
    \end{split}
\end{equation}
\noindent where $D_{j_1,j_2}$ is the number of subspaces $ V_{j_1, \ell_1, j_2, \ell_2}$ associated to the integers or half-integers $j_1$ and $j_2$ :
\begin{equation}
    \begin{split}
         D_{j_1, j_2} =  \frac{(2j_1 + 1)(2j_2 + 1)}{(n-k+1)(k+1)}\binom{n-k+1}{\frac{n-k}{2} - j_1}\binom{k+1}{\frac{k}{2} - j_2}.
    \end{split}
    \label{degen}
\end{equation}

Finally, we want to show that the subspaces $V_{j_1, j_2}$ are irreducible $\mathcal{T}$-submodules. The overlaps between the vectors in an irreducible representation of $\mathfrak{su}(2)$ $\ket{j, m}$ and the vectors in the basis yielded by the tensor product of two irreducible representations $\ket{j_1, m_1} \otimes \ket{j_2, m_2} =\ket{j_1, m_1, j_2, m_2}$ are the Clebsh-Gordan coefficents $c^{j,j_1,j_2}_{m,m_1,m_2}$ of $\mathfrak{su}(2)$. These are known to be given in terms of the dual Hahn polynomials $R_i(\lambda(x),\gamma, \delta, N)$ \cite{koornwinder1981clebsch}. Indeed, if $j_1 < j_2$ we have \cite{koornwinder1981clebsch}:
\begin{equation}
    \begin{split}
      c^{j,j_1,j_2}_{m,m_1,m_2}  = \bra{j \ m }\ket{j_1, m_1, j_2, m_2} &= \mathcal{N}\  R_i(x(x+\delta + \gamma + 1),\gamma,\delta, N),
    \end{split}
    \label{CG}
\end{equation}
\noindent with
\begin{align}
   \mathcal{N} = (-1)^i \sqrt{\frac{(N!)(-N)_x (\gamma + 1)_x (2x + \delta + \gamma + 1)}{(-1)^x(x!)(\delta + 1)_x(x+ \delta + \gamma + 1)_{N+ 1}}}\sqrt{\binom{\gamma + i}{i} }\sqrt{\binom{N  + \delta -i}{N-i}}
   \label{weig}
\end{align}
\noindent and
\begin{equation}
    \begin{split}
        & i = j_1 - m_1, \quad x = j_1 - j_2 + j, \quad N = 2j_1, \\ 
        &  \quad \ \  \delta = -j_1 + j_2 - m \quad \text{and} \quad \gamma = -j_1 + j_2 + m.
        \label{var}
    \end{split}
\end{equation}
\noindent For $j_1 > j_2$, one only needs to exchange $j_1$ with $j_2$ and $m_1$ with $m_2$ in \eqref{var}. From the three terms recurrence relation and difference equation of the dual Hahn polynomials, we find that the action of $A$ on the eigenvectors of $A^*$ and the action of $A^*$ on the eigenvectors of $A$ are irreducible tridiagonal. Thus,
\begin{align}
    \mathbb{C}^{|X|} = \bigoplus_{j_1, \ell_1, j_2, \ell_2} V_{j_1, \ell_1, j_2, \ell_2} = \bigoplus_{j_1, j_2} D_{j_1,j_2} V_{j_1, j_2}
    \label{deco}
\end{align}
\noindent corresponds to the decomposition in the irreducible $\mathcal{T}$-submodules we were looking for.
\subsection{The entanglement entropy for a single neighborhood}
\label{entents}
\begin{figure}[h]
\begin{subfigure}{.5\textwidth}
  \centering
  \includegraphics[scale = 0.95]{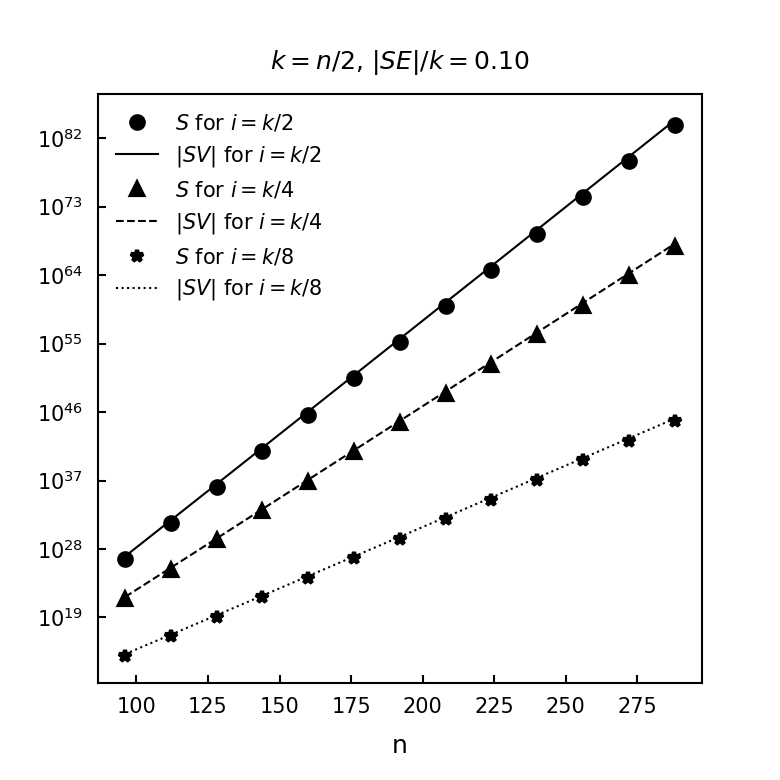}
  \caption{}
  \label{fig:sfig1}
\end{subfigure}%
\begin{subfigure}{.5\textwidth}
  \centering
  \includegraphics[scale =0.95]{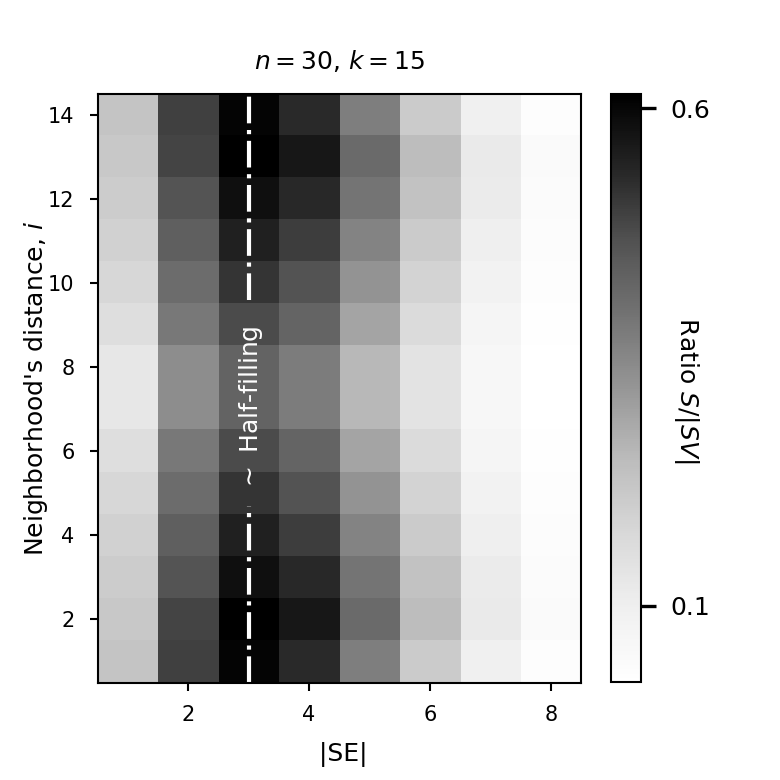}
  \caption{}
  \label{fig:sfig2}
\end{subfigure}
\caption{Entanglement entropy for single neighborhoods in fermionic systems on $J(n,n/2)$. (a): von Neumann entropy of the neighborhoods $i = k/2$, $i = k/4$ and $i = k/8$ when the first $1/10$ of the energy levels are occupied ($\sim$ half-filling). (b) Ratio of the entropy over the dimension of the subsystem in the case $n = 30$ and $k = 15$ for different neighborhood's distance $i$ and number of energy level occupied $|SE|$. }
\label{fig:fig2}
\end{figure}

We shall see that \eqref{deco} simplifies the computation of the entanglement entropy. Let us denote $\ket{j_1, m_1}_{\ell_1}\otimes\ket{j_2, m_2}_{\ell_2} = \ket{j_1, m_1, j_2, m_2}_{\ell_1, \ell_2}$ and define
\begin{align}
    \underbar{c}^\dagger_{\substack{j_1, m_1, \ell_1, \\ j_2, m_2, \ell_2}} = \sum_{x \in  X } \bra{x}\ket{j_1, m_1, j_2, m_2}_{\ell_1, \ell_2} c_{x}^\dagger
\end{align}
\noindent and
\begin{align}
    \underbar{c}_{\substack{j_1, m_1, \ell_1, \\ j_2, m_2, \ell_2}} = \sum_{x \in  X  } { }_{\ell_1, \ell_2}\langle j_1, m_1, j_2, m_2\ket{x} c_{x},
\end{align}
\noindent where we recall that $m_2 = n/2 - k - m_1$. These operators respect the canonical relations of fermionic creation and annihilation operators and allow to rewrite the Hamiltonian as
\begin{equation}
    \begin{split}
          \widehat{\mathcal{H}}  = \sum_{\substack{j_1,\ell_1, \\ j_2,\ell_2}} [\widehat{\mathcal{H}}]_{V_{j_1,\ell_1,j_2,\ell_2}},
    \end{split}
\end{equation}
\noindent where $[\widehat{\mathcal{H}}]_{V_{j_1,\ell_1,j_2,\ell_2}} = [\widehat{\mathcal{H}}]_{V_{j_1,j_2}}$ is given by
\begin{equation}
    \begin{split}
        \sum_{m_1,m_1'} {}_{\ell_1, \ell_2}\langle j_1,m_1,j_2,m_2|\Big[ \sum_{i = 0}^{k} \alpha_i A_i \Big]  \ket{j_1,m_1',j_2,m_2'}_{\ell_1, \ell_2} \underbar{c}^\dagger_{\substack{j_1, m_1, \ell_1, \\ j_2, m_2, \ell_2}} \underbar{c}^\dagger_{\substack{j_1, m_1', \ell_1, \\ j_2, m_2', \ell_2}}.
    \end{split}
    \label{Hamilchain}
\end{equation}
\noindent Since there is at most one vector per neighborhood of $x_0$ in a given submodule $V_{j_1,j_2}$, the operator \eqref{Hamilchain} can be understood as the Hamiltonian of free fermions hopping on a chain. Similarly, we find for the chopped correlation matrix that
\begin{equation}
    \begin{split}
    C &= \pi_{SV} \pi_{SE} \pi_{SV}\\
    &= \sum_{\substack{j_1,\ell_1, \\ j_2,\ell_2}} [\pi_{SV} \pi_{SE} \pi_{SV}]_{V_{j_1, \ell_1, j_2,\ell_2}} \\
    &= \sum_{\substack{j_1,\ell_1, \\ j_2,\ell_2}} [C]_{V_{j_1, \ell_1, j_2,\ell_2}} \ ,
    \end{split}
    \label{dec}
\end{equation}
\noindent where $[C]_{V_{j_1, \ell_1, j_2,\ell_2}} = [C]_{V_{j_1, j_2}}$ is the restriction of the chopped correlation matrix to a subspace ${V_{j_1, j_2}}$. Its entries are
\begin{equation}
    \begin{split}
             \langle j_1, m_1, j_2, m_2|[C]_{V_{j_1,j_2} }\ket{j_1, m_1',j_2, m_2'} &= \sum_{j \in SE}c^{j,j_1,j_2}_{m_1 + m_2 ,m_1,m_2} c^{j,j_1,j_2}_{m_1' + m_2' ,m_1',m_2'},
    \end{split}
\end{equation}
\noindent with the Clebsh-Gordan coefficients $c_{m_1 + m_2, m_1, m_2}^{j,j_1,j_2}$ being given by \eqref{CG}. While the dimension of the chopped correlation matrix is equal to the number of sites in all the neighborhoods of the subsystem, the dimension of its submatrices is 
\begin{align}
    \text{dim}[C]_{V_{j_1, j_2}} = \#\{i \in SD : j_1 \geq |(n-k)/2 - i|,j_2 \geq |k/2 - i| \}.
\end{align}
 This is at most the number of distances in $SD$. The decomposition thus significantly simplify the diagonalization of the chopped correlation matrix. In fact, if the subsystem is made out of a single neighborhood, the decomposition yields the eigenvalues of $C$. Indeed, the submatrices are then at most $1-$dimensional and their unique entry is an eigenvalue of $C$. If the subsystem 1 is the $i^{\text{th}}$ neighborhood of $x_0$, we have:
\begin{align}
    \lambda_{j_1,j_2} = \sum_{j \in SE}\left(c^{j,j_1,j_2}_{m_1 + m_2 ,(n-k)/2 - i,i - k/2}\right)^2.
    \label{eigen}
\end{align}
\noindent The degeneracy $D_{j_1,j_2}$ of $\lambda_{j_1,j_2}$ is given by \eqref{degen}. These formulas can be used as inputs in \eqref{entropy} to obtain an analytical expression for the von Neumann entropy. Figure \ref{fig:fig2} presents results obtained for different values of $n$, of neighborhood's distance $i$ and of number of energy levels occupied $|SE|$. 

On the left figure, we see that the entanglement entropy is bounded by the number of sites in $SV$. Since all the vertices in a neighborhood are on the boundary of the subsystem, this is consistent with an area law. 

The figure on the right shows that the ratio of the entanglement entropy over the number of sites in $SV$, i.e. the pre-factor of the area law, peaks when a small fraction of the energy levels are occupied. This is due to the massive degeneracy $D_j$ of the lowest energy levels. For instance, while free fermions on $J(30,15)$ have $k+1 = 16$ energy levels, approximately half of the single-particle excitations are associated to the first three of them. Looking at the system in terms of its path decomposition \eqref{deco}, this is translated as the absence of high energy levels in most chains. Finally, let us  note that the symmetry of the right figure with respect to $i = 7.5$ is due to the equivalence of the neighborhoods $i$ and $k-i$ when $k = n/2$. 

\subsection{$\mathcal{T}$ and the Hahn algebra}

\noindent Before considering subsystems composed of multiple neighborhoods, we shall make an additional remark concerning $\mathcal{T}$. We recall that \eqref{Adjf} and \eqref{Adjff} give an expression for the generators of the Terwilliger algebra of the Johnson scheme in terms of generators from two copies of $\mathfrak{su}(2)$. We can use this connection to compute the commutation relations for pairs of elements in $\mathcal{T}$. Let 
\begin{align}
    \hat{K}_1 = \frac{2k(n-k)}{n(n-1)} A^* , \quad \quad \hat{K}_2 = A
\end{align}
 \noindent and take $\textbf{j}^2_1$ and $\textbf{j}^2_2$ to be the Casimir operators associated to the first and second copy of $\mathfrak{su}(2)$. Note that these are central elements in $\mathcal{T}$. One can check that the following relations are verified:
\begin{align}
    [\hat{K}_1, \hat{K}_2] = \hat{K}_3,
    \label{H1}
\end{align}
\begin{align}
    [\hat{K}_2, \hat{K}_3] = a \{\hat{K}_1, \hat{K}_2\} + b \hat{K}_2 + c_1 \hat{K}_1 + d_1
    \label{H2}
\end{align}
\noindent and
\begin{align}
    [\hat{K}_3, \hat{K}_1] = a \hat{K}_1^2 + b\hat{K}_1 + c_2 \hat{K}_2 + d_2,
    \label{H3}
\end{align}
\noindent where
\begin{equation}
    \begin{split}
        & a = -2, \quad b = -\frac{2(n-2k)^2}{n},\\
        &  c_1 = -(n-2k) - 2n, \quad c_2 = -4,\\
        & d_1 = -\frac{b c_1}{4} + (n-2k)(\textbf{j}_1^2 - \textbf{j}_2^2),\\  & d_2 = -2n  + 4(\textbf{j}^2_1 + \textbf{j}^2_2) - \frac{b^2}{8}  + \frac{bn}{4}.
    \end{split}
\end{equation}
\noindent Since \eqref{H1}, \eqref{H2} and \eqref{H3} are the defining relations of the Hahn algebra $\mathfrak{h}$, we find that the Terwilliger algebra of the Johnson scheme is a quotient of a central extension of $\mathfrak{h}$. The Hahn algebra was introduced in \cite{GRANOVSKII19921} to describe the symmetry properties of the dual Hahn polynomials, which appeared in equation \eqref{Hahn}, in equation \eqref{CG} and are related to the Johnson scheme, as already pointed out at the beginning of section \ref{s3}.

\section{The generalized algebraic Heun operator}
\label{s4}
\begin{figure}[h]
\begin{subfigure}{.5\textwidth}
  \centering
  \includegraphics[scale = 0.95]{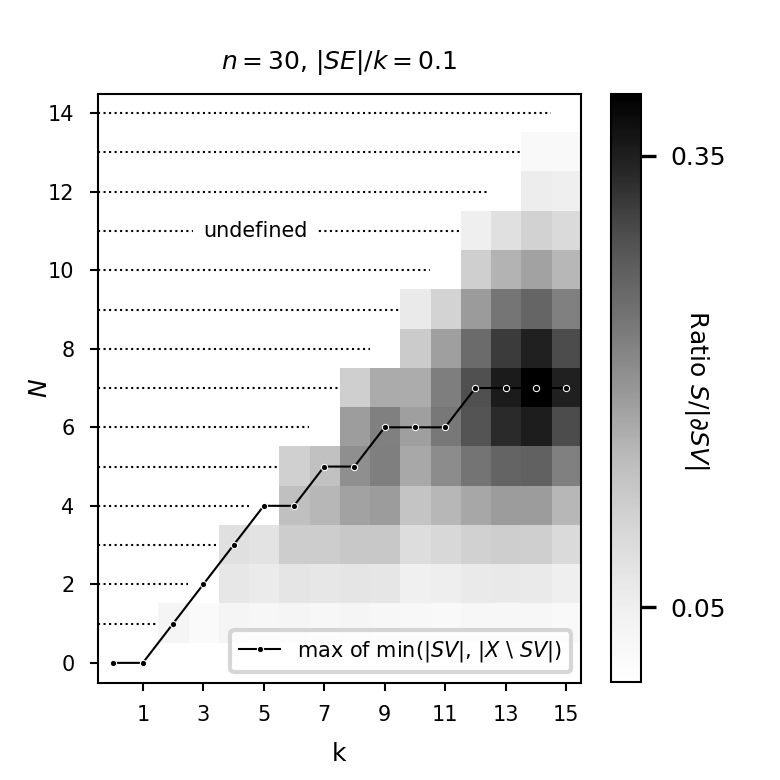}
  \caption{}
  \label{fig:sfig31}
\end{subfigure}%
\begin{subfigure}{.5\textwidth}
  \centering
  \includegraphics[scale = 0.95]{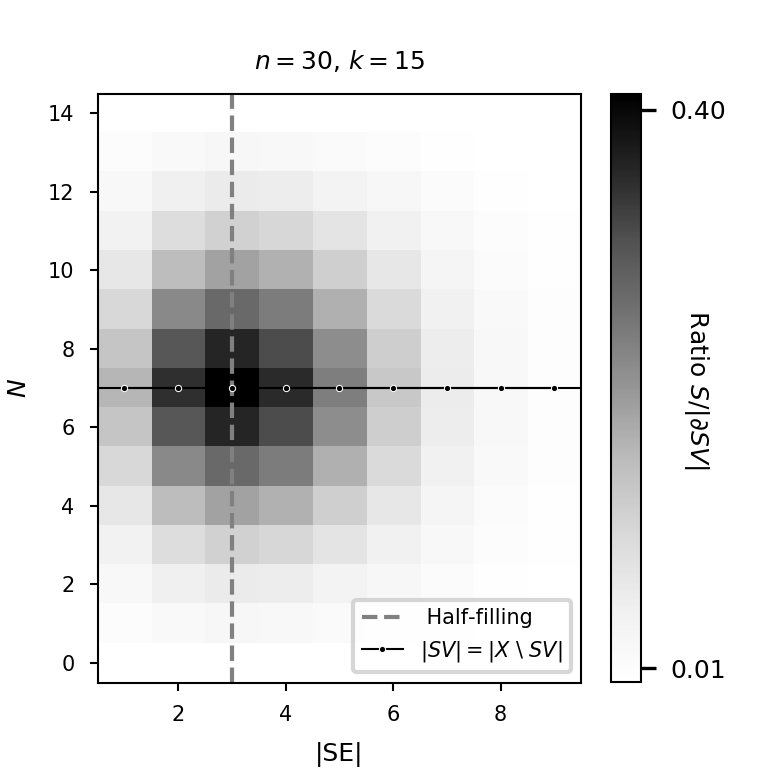}
  \caption{}
  \label{fig:sfig32}
\end{subfigure}

\caption{Ratio of the von Neumann entropy $S$ over the size of the boundary $\partial SV$ of the subsystem for free fermions on $J(30,k)$. (a) Ratio $S/|\partial SV|$ for different diameter $k$ of the Johnson graph and for different number of neighborhoods in $SV$. (b) Ratio $S/|\partial SV|$ for a different number of energy levels filled and for different number of neighborhoods in $SV$.  }
\label{fig:fig3}
\end{figure}

\noindent In this section, we are interested in the case where $SV$ is the set of sites at a distance lower than some large integer $N+1$ from a given vertex, i.e.
\begin{align}
    \pi_{SV} = \sum_{i = 0}^N E_i^*.
\end{align}
\noindent The submatrices in formula \eqref{dec} have a dimension proportional to the number of neighborhoods in $SV$, have mostly non-zero entries and have many eigenvalues near 0 and 1. Consequently, diagonalizing the operators $[C]_{V_{j_1, j_2}}$ is less practical as $N$ increases and we shall turn to an approach developed to study time and band limiting problems \cite{ Cramp__2019, Cramp__2020,Landau1985,Slepian1983}. We thus look for an operator $T$ with practical diagonalization properties that verify
\begin{align}
    [C, T] = 0.
    \label{commutebaby}
\end{align}
\noindent Such an operator would share with $C$ common eigenvectors. Recently, the diagonalization of the chopped correlation matrix associated to free fermions on general distance-regular graphs was considered and a way of constructing $T$ was presented \cite{crampe2020entanglement}. For Johnson graphs, it requires to look at the most general symmetric block-tridiagonal operator in $\mathcal{T}$:
\begin{align}
    T = \{A, A^*\} + \mu A^* + \nu A.
\end{align}
\noindent $T$ is referred to as a generalized algebraic Heun operator \cite{GB_2018}. We want to fix $\mu$ and $\nu$ so that $T$ commutes with both $\pi_{SV}$ and $\pi_{SE}$, assuring that it also commutes with $C$. The action of the adjacency matrix and dual adjacency matrix is obviously diagonal on their own eigenbasis. Moreover, one can check that the action of $A$ on the eigenbasis of $A^*$ is tridiagonal:
\begin{equation}
    \begin{split}
        A \ket{j_1,m_1,j_2,m_2} & = a_{m_1+1} \ket{j_1,m_1+1,j_2,m_2-1} + b_{m_1}\ket{j_1,m_1,j_2,m_2} \\
        &\quad + a_{m_1} \ket{j_1,m_1-1,j_2,m_2+1}
    \end{split}
\end{equation}
\noindent where
\begin{align}
    a_{m_1} = \sqrt{(j_1 + m_1)(j_1 - m_1 + 1)(j_2 - m_2)(j_2 + m_2 + 1)}
\end{align}
\noindent and
\begin{align}
    b_{m_1} = (j_1+1)(j_1) + (j_2+1)(j_2) - m_1^2 - m_2^2 - \frac{n}{2}.
\end{align}
\noindent The same is true for the action of $A^*$ on the eigenvectors of $A$. Let us denote $c^{j,j_1,j_2}_{m,m_1,m_2}= c^j_{m_1}$. We find that 
\begin{equation}
    \begin{split}
            A^* \ket{j, n/2 - k} &= A^* \sum_{m_1} c^{j}_{m_1} \ket{ j_1, m_1, j_2,n/2 - k - m_1} \\
            &= \sum_{m_1}( a^*_{j+1} c^{j+1}_{m_1} + b^*_j c^j_{m_1} + a^*_j c^{j-1}_{m_1}) \ket{ j_1, m_1, j_2,n/2-k - m_1}\\
            & = a^*_{j+1} \ket{j+1\  n/2 - k} + b^*_j \ket{j\  n/2 - k} + a^*_j \ket{j-1 \  n/2 - k},
    \end{split}
\end{equation}
\noindent where
\begin{align}
    a^*_j = \frac{n(n-1)}{k(n-k)}\left(\frac{(j^2 - m^2)(j^2 - (j_1 - j_2)^2)((j_1 + j_2+1)^2 - j^2)}{(4j^2 -1)(4j^2)}\right)^{\frac{1}{2}}
\end{align}
\noindent and
\begin{align}
    b^*_j =  -\frac{(n-1)(n-2k)}{2k} + \frac{n (n-1)(n-2k) }{2k(n-k)}\left(\frac{1}{2} + \frac{(j_1 + j_2 + 1)(j_1 - j_2)}{2j(j+1)}\right).
\end{align}
\noindent We can use these formulas to express the action of the generalized algebraic Heun operator on both basis. We find that
\begin{equation}
    \begin{split}
          T \ket{j, m } &= a^*_{j+1} (\theta_{j+1} + \theta_{j} + \mu) \ket{j+1,  m}  \\&\quad+ (\mu b_j^* + \nu \theta_j + 2 b_j^*\theta_{j})\ \ket{j, m}+  a^*_{j} (\theta_{j-1} + \theta_{j} + \mu) \ket{j-1, m} 
    \end{split}
    \label{T1}
\end{equation}
\noindent and
\begin{equation}
    \begin{split}
          T \ket{j_1,  m_1,j_2,  m_2 } &= a_{m_1+1} (\theta^*_{m_1 +1, m_2 - 1} +\theta^*_{m_1, m_2} + \nu) \ket{j_1, m_1 + 1,j_2, m_2-1 } \\&\quad+ (\nu b_{m_1} +  \mu \theta^*_{m_1, m_2} + 2 b_{m_1}\theta^*_{m_1, m_2}) \ket{j_1, m_1,j_2, m_2}\\
          & \quad +   a_{m_1} ( \theta^*_{m_1 -1, m_2 + 1} +\theta^*_{m_1, m_2}+ \nu)  \ket{j_1,  m_1 - 1,j_2, m_2+1 }.
          \label{T2}
    \end{split}
\end{equation}
\noindent The restriction of this operator $T$ to a single module $V_{j_1,j_2}$ corresponds to an affine transformation of the Heun operator constructed to study the dual Hahn fermionic chain in \cite{Cramp__2019}. In the case where $SV = \{x \in X : d(x_0,x) \leq N\}$ and $SE = \{\frac{n}{2}-k, \dots, j_0\}$, the action of the projectors $\pi_{SV}$ and $\pi_{SE}$ is given by
\begin{align}
\pi_{SV} \ket{j_1,m_1,j_2,m_2} = 
         \left\{
    \begin{array}{ll}
    	 \ket{j_1,m_1,j_2,m_2} & \mbox{if } \frac{n-k}{2} - m_1 \leq N,\\
    	0 & \mbox{otherwise }
    \end{array}
\right. 
\label{P1}
\end{align}
\noindent and
\begin{align}
\pi_{SE} \ket{j, n/2 - k} = 
         \left\{
    \begin{array}{ll}
    	 \ket{j,n/2 - k} & \mbox{if } j \leq j_0,\\
    	0 & \mbox{otherwise. }
    \end{array}
\right. 
\label{P2}
\end{align}
\noindent One can check using \eqref{T2} and \eqref{P1} that $[T, \pi_{SV}] = 0$ if
\begin{align}
     \theta^*_{\frac{n-k}{2} - N - 1 , -\frac{k}{2} + N + 1} +\theta^*_{\frac{n-k}{2} - N , -\frac{k}{2} + N} + \nu = 0. 
\end{align}
\noindent Similarly, we find with \eqref{T1} and \eqref{P2} that $[T, \pi_{SE}] = 0$ if
\begin{align}
    \theta_{j_0+1} + \theta_{j_0} + \mu = 0. 
\end{align}
\noindent In other words, condition \eqref{commutebaby} is verified if we choose
\begin{align}
    \mu = -\theta_{j_0+1} - \theta_{j_0} \quad \text{and} \quad \nu = -\theta^*_{\frac{n-k}{2} - N - 1 , -\frac{k}{2} + N + 1}  -\theta^*_{\frac{n-k}{2} - N , -\frac{k}{2} + N}.
\end{align}
\noindent Since $T$ is tridiagonal on each module $V_{j_1,j_2}$, it is in general easier to diagonalize than $C$. Once we have its eigenvectors, we can act on them with the chopped correlation matrix and read out the spectrum of $C$ from the outcome. Figure \ref{fig:fig3} presents von Neumann entropies associated to subsystems $SV = \{x \in X  : d(x_0,x) \leq N\}$ of $J(30,k)$ that were obtained by this method and by diagonalizing $T$ numerically. 

As expected, the entanglement entropy is bounded by the number of sites on the boundary $\partial SV$, i.e. the number of sites in the $N^{\text{th}}$ neighborhood. We also see that the pre-factor $S/|\partial SV|$ of the area law peaks when only the lowest energy levels are occupied and when the number of sites in both the subsystem $SV$ and its complement $X \backslash SV$ is large. The first condition was discussed in subsection \ref{entents} and has to do with the important degeneracy of these levels. The second condition suggests that the area law pre-factor for this system depends more on the entanglement of the bulk than on the entanglement of the boundary. Indeed, figure \ref{fig:sfig31} shows that the ratio $S/|\partial SV|$ reaches its maximum value when the subsystem $SV$ and its complement $X \backslash SV$ are both large. In figure \ref{fig:sfig32} which represents the symmetric case $k = n/2$, this happens at $N \approx k/2$. By contrast, the ratio of the entanglement entropy of the $N^{\text{th}}$ neighborhood (the boundary between the two regions) over its size is at its lowest when $N \approx k/2$. This is shown in figure \ref{fig:sfig2}.

\begin{figure}[h]
\begin{subfigure}{.5\textwidth}
  \centering
  \includegraphics[scale = 1]{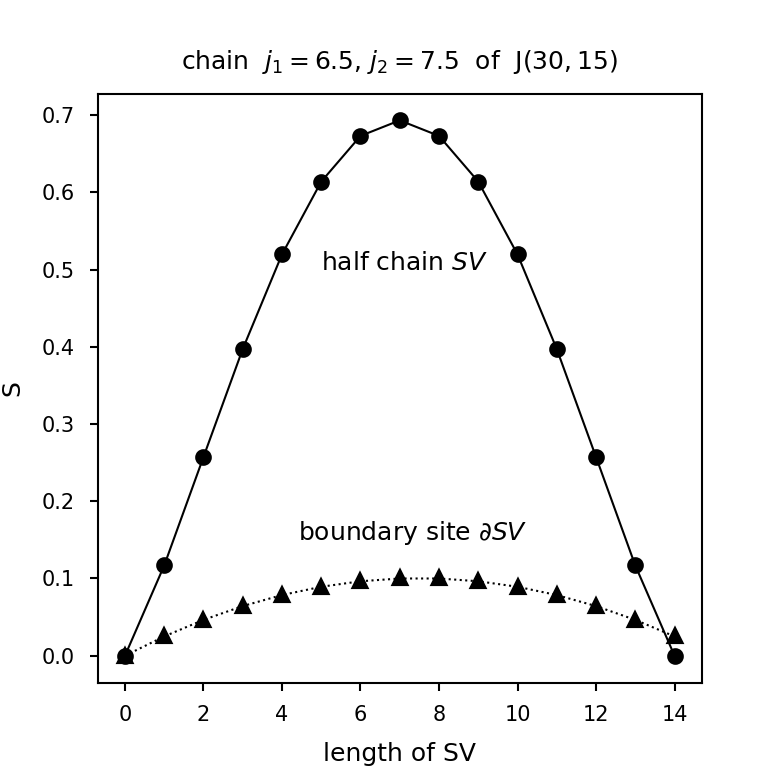}
  \caption{}
  \label{fig:sfig41}
\end{subfigure}%
\begin{subfigure}{.5\textwidth}
  \centering
  \includegraphics[scale = 1]{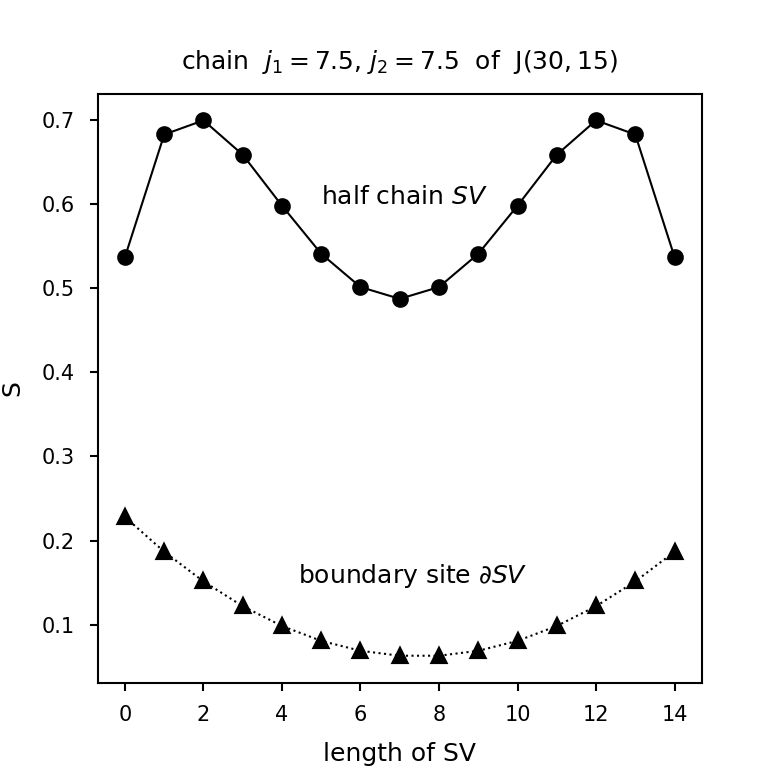}
  \caption{}
  \label{fig:sfig42}
\end{subfigure}

\caption{Entranglement entropy of half chains, for chains in $J(30,15)$.  (a) Entanglement entropy of half chains $SV$ of different lengths and of their boundary site $\partial SV$ in the case $j_1 = 6.5$, $j_2 = 7.5$. (b) Entanglement entropy of half chains $SV$ of different lengths and of their boundary site $\partial SV$ in the case $j_1 = j_2 = 7.5$.}
\label{fig:fig4}
\end{figure}

\noindent It should be stressed that, even if Johnson graphs are equivalent to bundles of chains, a model of free fermions on $J(n,k)$ does not share all the properties of one dimensional systems. For instance, the dominating role of the bulk in the pre-factor only emerges when considering the graph. It does not appear for an individual chain in the path decomposition \eqref{deco} of $J(n,k)$, where the entanglement of a region behaves similarly to the entanglement of its boundary site (see figure \ref{fig:fig4}). Since the relation between the entanglement entropy $S$ and the length of a half chain changes for different values of $j_1$ and $j_2$, it seems that the correlation between the entanglement of a region and the entanglement of its shell is lost when one looks at the graph and sums over the contribution of each path. 

\section{Concluding remarks}
We have investigated the entanglement entropy of sets of neighborhoods in systems of free fermions living on the vertices of Johnson graphs. For a subsystem composed of a single neighborhood, we have provided an analytical expression. It was given by the decomposition in irreducible representations of the Terwilliger algebra of the Johnson scheme, which was obtained by embedding the algebra in two copies of the Terwilliger algebra of the hypercube. It was also shown that $\mathcal{T}$ is a quotient of the centrally extended Hahn algebra $\mathfrak{h}$. For subsystems composed of many neighborhoods, we have constructed a simple block-tridiagonal operator $T$ commuting with the chopped correlation matrix. As a difference analog of a second order differential operator, it is prone to possess a well-distributed spectrum which allows to compute easily the entanglement entropy numerically for large systems and to investigate area law pre-factors. 

A similar approach was used in \cite{Hamming} to study the entanglement entropy of free fermions on Hamming graphs. In that case, the block-tridiagonal operator $T$ was a Heun operator of Lie type that could be diagonalized analytically by algebraic Bethe ansatz methods \cite{ bernard2020heun,Crampe_2017}. Since the FRT presentation of the Hahn algebra is known \cite{Cramp__2019trunc}, we expect that the algebraic Bethe ansatz could also be used to diagonalize the operator $T$ of Hahn type which arise for Johnson graphs. Future work could be oriented in this direction. Moreover, it should prove interesting to study the entanglement of free fermions on other distance-regular structures, such as the dual polar and Grassman graphs \cite{brouwer2012distance} which are related to $q$-polynomials of the Askey-scheme \cite{thiniii}.
\section*{Acknowledgements}

We thank Krystal Guo for discussions. PAB holds a scholarship from the Natural Sciences and Engineering Research Council of Canada (NSERC). The research of LV is supported in part by a Discovery Grant from NSERC.

\end{document}